\documentclass[aps,prl,nobibnotes,twocolumn,superscriptaddress,bibliography]{revtex4-2}
\usepackage{amsfonts}
\usepackage{mathrsfs}
\usepackage{amsmath}
\usepackage{color}
\usepackage{natbib}
\usepackage{graphicx}
\usepackage{bm}
\usepackage{amssymb}
\usepackage{xspace}
\usepackage{epstopdf}
\usepackage{dcolumn}
\usepackage{longtable}
\usepackage{array}
\usepackage{makecell}
\usepackage{tabulary}
\usepackage{multirow}
\usepackage{appendix}
\usepackage{mathtools}
\usepackage{pifont}
\usepackage{longtable}
\newcolumntype{C}[1]{>{\centering\arraybackslash}p{#1}}
\newcolumntype{L}[1]{>{\flushleft\arraybackslash}p{#1}}

\usepackage{multirow}
\usepackage{epsfig}
\usepackage[normalem]{ulem}
\usepackage{amsmath}
\usepackage{braket}
\usepackage{bbold}
\usepackage{graphicx}

\makeatletter

\usepackage[colorlinks=true, letterpaper=true, pdfstartview=FitV, linkcolor=blue, citecolor=blue, urlcolor=blue]{hyperref}

\makeatother
\begin{document}


\title{Multiple Topological Phases Controlled via Strain in Two-Dimensional Altermagnets}

\author{Zesen Fu}
\thanks{These authors contributed equally to this work.}
\affiliation{School of Physics and Technology, Xinjiang University, Urumqi 830017, China}
\affiliation{School of Physics, Central South University, Changsha 410083, China}

\author{Mengli Hu}
\thanks{These authors contributed equally to this work.}
\affiliation{Leibniz Institute for Solid State and Materials Research, IFW Dresden, Helmholtzstraße 20, 01069 Dresden, Germany}

\author{Aolin Li}
\email{liaolin628@xju.edu.cn}
\affiliation{School of Physics and Technology, Xinjiang University, Urumqi 830017, China}

\author{Haiming Duan}
\affiliation{School of Physics and Technology, Xinjiang University, Urumqi 830017, China}

\author{Junwei Liu}
\affiliation{Department of Physics, The Hong Kong University of Science and Technology, Hong Kong, People's Republic of China}

\author{Fangping Ouyang}
\email{ouyangfp06@tsinghua.org.cn}
\affiliation{School of Physics and Technology, Xinjiang University, Urumqi 830017, China}
\affiliation{School of Physics, Central South University, Changsha 410083, China}
\date{\today}

\begin{abstract}
Altermagnets (AMs) are an emergent class of magnetic materials that combine properties of ferromagnets and antiferromagnets, exhibiting spin-polarized Fermi surfaces and zero net magnetic moment due to combined time-reversal and crystal symmetry. Here, we construct a Kondo-lattice model on a two-dimensional square Lieb lattice to investigate the topological properties of AMs. We identify a type-II quantum spin Hall state characterized by spin-polarized counterpropagating edge states. Breaking the $C_{4z}\mathcal{T}$ symmetry, which connects magnetic sublattices, induces a transition to a quantum anomalous Hall state. We further establish a strain-induced mechanism to control these topological phase transitions and present the corresponding phase diagram. Finally, we demonstrate the predicted transitions in monolayer CrO, a realistic altermagnetic candidate, using first-principles calculations. Our findings highlight the potential of 2D AMs as a versatile platform for topological spintronics, enabling strain-tunable helical and chiral edge states within a single system.

\end{abstract}


\maketitle


\textit{Introduction.}---Topological insulators (TIs) are characterized by topologically protected conducting edge states co-existing with the insulating bulk. In two dimensions, the quantum spin Hall (QSH) effect \cite{RN7,RN8,RN9,RN10} and the quantum anomalous Hall (QAH) effect \cite{2010rmpM,2011rmpX,2006prbX} manifest their nontrivial topology through helical and chiral edge states, protected by the $\mathbb{Z}_2$ invariant and the Chern number, respectively. These dissipationless edge states have attracted significant attention for their potential in low-power electronic devices. While several material candidates have been identified, the ability to engineer topological properties, such as controlling the Chern number or inducing phase transitions, remains a central challenge. Established approaches include applying external magnetic fields, magnetic doping \cite{2013sJ,2013sC,2010sR,2014npJ,2014prlX,2015nmC,2018amO}, using magnetic substrates \cite{2020nZ,2022prbZ,2020sY,2020ncmP,2017saC,2020saH,2020prbZ}, or reorienting magnetization in ferromagnets \cite{2022prlL,2023nsrX}. However, switching the material between QSH and QAH states remains challenging due to the fundamental distinction imposed by time-reversal symmetry, making this problem scientifically intriguing.

Altermagnets, a recently discovered class of materials, offer a promising alternative by combining properties of both ferro- and antiferromagnets \cite{2022prx1,2022prx2,2022prx3,2021ncH}. These materials exhibit spin-split bands and alternating spin-polarized Fermi surfaces, arising from their unique spin-lattice symmetry \cite{2022prx1,2022prx2,2022prx3,2021ncH,2014japC,2019jpsjS,2020prbL,2021prlR,2021ncH}. The breaking of time-reversal times inversion symmetry ($\mathcal{T}$) enforces non-degenerate bands, and the net zero magnetic moment is guaranteed by time-reversal times crystal symmetry--altermagnetic symmetry--connecting the magnetic sublattices~\cite{2022prx1,2022prx2,2022prx3,2021ncH}. This altermagnetic symmetry enables spin-valley locking and even-wave Fermi surfaces in momentum space \cite{2024prlX,2021ncH}, with equivalent spin-up and spin-down populations. 
Altermagnets have been predicted to host a variety of unconventional phenomena, including large spin-polarized currents \cite{2021ncH,RN18,2021prlR,RN26,RN27,RN28,RN29,RN30,RN45,2025prlC,2025prlL}, piezomagnetism \cite{2021ncH,RN18,RN27}, the anomalous Hall effect \cite{2020saL,2021prlR,2019ncM,RN28,2024prbO,RN34,2024ncR,2020prbN,2024prlL}, Weyl altermagnetism \cite{RN45,AlterWeyl}, and spin-polarized edge states \cite{2025aF}. Some of these phenomena have already been observed experimentally \cite{RN27,RN28,RN29,RN30,RN32,RN41,RN42,2019ncM,RN34,2024prbO,2024ncR,2024prlL}, while their interplay with topological phases remains an open question.

In this work, we develop a minimal 2D Kondo-lattice model to capture the essential altermagnetic properties. Band inversions occur around two valleys with opposite spins, leading to a type-II quantum spin Hall (QSH) state \cite{2025aF}, where spin-polarized edge modes propagate in opposite directions, protected by collinear magnetic order and altermagnetic symmetry ($C_{4z}\mathcal{T}$). By breaking $C_{4z}\mathcal{T}$, the system can transition to quantum anomalous Hall (QAH) states with opposite Chern numbers. We further propose biaxial and uniaxial strains as effective tools to realize and control these topological phases. Finally, we investigate a representative altermagnetic candidate, monolayer CrO, and demonstrate the predicted topological phase transitions using first-principles calculations. The paper is organized as follows: we first introduce the model with and without symmetry breaking, then discuss the topological phase diagram and its criticalities, and finally present first-principles results on monolayer CrO.

\begin{figure}
	\centering 
	\includegraphics[width=0.5\textwidth]{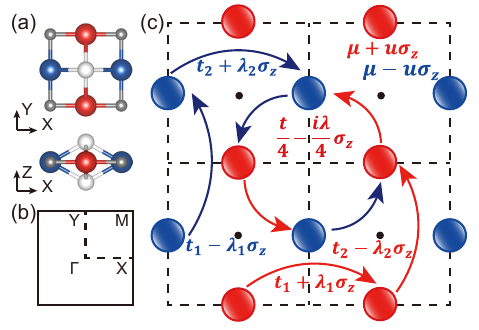}%
	\caption{\label{fig:1} 
		(a) Top view and front view of the lattice structure. Red and blue spheres represent the same magnetic atoms with opposite magnetization, white and gray spheres correspond to different nonmagnetic atoms.
		(b) Schematic diagram of the first Brillouin zone.
		(c) Schematic structure of the altermagnet.
		Red and blue atoms denote magnetic ions with opposite magnetizations, whose onsite potentials are $\mu + u\sigma_z$ and $\mu - u\sigma_z$, respectively.
		The arrows indicate NN and NNN hoppings, while the counterclockwise NN hopping amplitudes are $t/4 - i\lambda \sigma_z/4$.}
\end{figure}

\textit{Model.}---We construct a two-dimensional altermagnetic model based on a square Lieb lattice, as illustrated in Fig.~\ref{fig:1}(c). The red and blue atoms represent magnetic sublattices with opposite spins, while the white and gray atoms in Fig.~\ref{fig:1}(a) denote nonmagnetic atoms that break the time-reversal times translation symmetry. Candidate materials for this model include CrO~\cite{2023aplXC}, Nb$_2$SeTeO~\cite{2025aF}, and V$_2$Se$_2$O~\cite{2021ncH}. The model incorporates essential symmetries, such as $C_{4z}\mathcal{T}$, which connects the two magnetic sublattices, $M_x\mathcal{T}$, the on-site symmetry of magnetic atoms, and $[U_{\bf{n}}(\pi)T||1]$, a unique symmetry in the spin space group that exists only in the absence of spin-orbit coupling.

Considering all symmetry constraints, we construct a four-band Kondo-lattice model with the basis $\ket{i,\sigma}$, where $i$ and $\sigma$ denote the sublattice and spin degrees of freedom, respectively. The nonmagnetic atoms are treated as an effective field that modifies the hopping parameters, ensuring that the Hamiltonian respects the symmetries described above. The model includes nearest-neighbor (NN) and next-nearest-neighbor (NNN) hoppings, off-site spin-orbit coupling (SOC), and collinear magnetic ordering described via the Kondo coupling to the itinerant electrons (see Supplemental Material~\cite{SM} for details).

\begin{widetext}
\begin{equation}
	\begin{aligned}
		\mathcal{H}(k_x,k_y) = & 
		\Bigl[\mu + A(\cos k_x + \cos k_y)\Bigr]\tau_0\sigma_0 + 
		B\Bigl[\cos k_x - \cos k_y\Bigr]\tau_z\sigma_0 + t\cos \frac{k_x}{2}\cos \frac{k_y}{2}\tau_x\sigma_0\\
		& + \lambda\sin \frac{k_x}{2}\sin \frac{k_y}{2}\tau_y\sigma_z + C\Bigl[\cos k_x - \cos k_y\Bigr]\tau_0\sigma_z+ \Bigl[u + D(\cos k_x + \cos k_y)\Bigr]\tau_z\sigma_z
	\end{aligned}
	\label{eq:1}
\end{equation}
\end{widetext}

Here, $\sigma_i$ and $\tau_i$ are the Pauli matrices representing the spin and sublattice degrees of freedom, respectively. The schematic plot of hoppings in the lattice is illustrated in Fig.~\ref{fig:1}(c), where $t$, $t_1$, and $t_2$ denote the hopping strengths, $\lambda$, $\lambda_1$, and $\lambda_2$ represent SOC, and $u$ characterizes the local magnetic moment. For convenience, we define $A = t_1 + t_2$ and $B = t_1 - t_2$, which describe the isotropic and anisotropic hoppings of NNNs, arising from the effective field of nonmagnetic atoms. Similarly, we define $C = \lambda_1 + \lambda_2$ and $D = \lambda_1 - \lambda_2$ for the NNN off-site SOC, which distinguishes this model from previous studies~\cite{RN45,2025aF}. Notably, this SOC term plays a crucial role in breaking particle-hole symmetry, which is more realistic for altermagnetic systems. The eigenvalues of the Hamiltonian are given by
\begin{equation}
	\begin{aligned}
		\epsilon_{\uparrow} &= E_f + s \pm \sqrt{v + \delta_{\uparrow}}, \\ 
		\epsilon_{\downarrow} &= E_f - s \pm \sqrt{v + \delta_{\downarrow}},
	\end{aligned}
\end{equation}
where $E_f = \mu + A(\cos k_x + \cos k_y)$, $s = C(\cos k_x - \cos k_y)$, $v = (t\cos \frac{k_x}{2}\cos \frac{k_y}{2})^2 + (\lambda\sin \frac{k_x}{2}\sin \frac{k_y}{2})^2$, and $\delta_{\uparrow/\downarrow} = [B(\cos k_x - \cos k_y) \pm (u + D(\cos k_x + \cos k_y))]^2$. Each eigenstate is spin-polarized due to the commutation relation $\left[\mathcal{H}, \tau_0 \otimes \sigma_z\right] = 0$, and we label the eigenvalues as $\epsilon_{\uparrow/\downarrow}$ for spin up/down. In the following discussion, $H_{\uparrow/\downarrow}$ is used to denote the Hamiltonian for the two spin species, satisfying $\mathcal{H} = H_{\uparrow} \otimes (\mathbf{I}_2 + \sigma_z)/2 + H_{\downarrow} \otimes (\mathbf{I}_2 - \sigma_z)/2$.



\begin{figure}
	\centering 
	\includegraphics[width=0.5\textwidth]{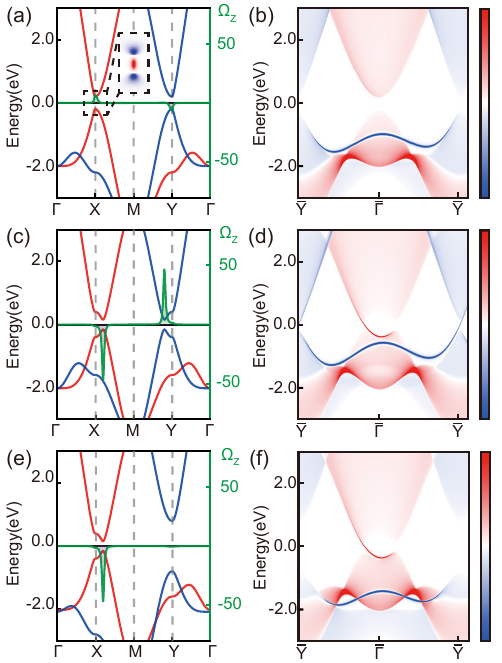}%
	\caption{\label{fig:2}Band structures and Berry curvature for (a) $u=-2.2$ and (c) $u=-1.6$. The other parameters are set as $A=0.5$, $B=-1$, $C=0.5$, $D=1$, $\lambda=0.5$, $\mu=1$, and $t=4$. The details of the Berry curvature near the X valley in the trivial phase are also shown in (a). The local density of states (LDOS) of the corresponding edge states are shown in (b) and (d). Here, Spin-polarized LDOS is obtained by subtracting the spin-down component from the spin-up component: deeper red indicates a larger contribution from spin-up states, while deeper blue corresponds to spin-down states. In (e) and (f), we explicitly break the $C_{4z}\mathcal{T}$ symmetry by adding an additional term $\mu_s\tau_z\sigma_0$($\mu_s=0.6$ and $u=-2.2$), and present the resulting band structure and edge states.}
	
\end{figure}

\textit{Topological Properties.}---To investigate the topological properties of the altermagnetic system, we fine-tune the model parameters to allow possible band inversion near the X and Y high-symmetry points. Due to the altermagnetic symmetry $C_{4z}\mathcal{T}$, the spin-up and spin-down bands align in energy at the X and Y points, respectively, as shown in Fig.~\ref{fig:2}(a). The energy gap across the Brillouin zone is given by
\begin{equation}
	\Delta_\mathrm{X/Y} = 2|u - 2B|,
	\label{eq:3}
\end{equation}
which corresponds to the gap values at the X and Y points. The critical condition for the topological phase transition is thus $u = 2B$. 

We compute the Berry curvature ($\Omega_z$) to further characterize the topological properties. Benefiting from the simple Hamiltonian, we derive the analytical expression of $\Omega_z$ for general momentum:
\begin{equation}
	\begin{aligned}
	\Omega_z(k_x,k_y)&=\Omega_z^{\uparrow}(k_x,k_y)+\Omega_z^{\downarrow}(k_x,k_y) \\
	& = \Omega_z^{\uparrow}(k_x,k_y)-\Omega_z^{\uparrow}(-k_y,k_x),\\
	\Omega_z^{\uparrow}(k_x,k_y)&=\frac{t\lambda}{16|v+\delta_{\uparrow}|^{\frac{3}{2}}} \Bigl[\sin^2{k_x}(B+D)+\sin^2{k_y}(B-D) \\
	& +2B(1-\cos{k_x}\cos{k_y})+u(\cos{k_x}-\cos{k_y})\Bigr],
	\label{berry}
	\end{aligned}
\end{equation}
where the NN hopping $t$ and SOC $\lambda$ are essential for nonzero $\Omega_z$. The sign change of $\Omega_z$ at X/Y is directly tied to the band inversion condition $u = 2B$ from Eq.~\eqref{berry}. As shown in Fig.~\ref{fig:2}(a) and (c), $\Omega_z$ is concentrated at around X and Y valleys and the opposite sign is constrained by $C_{4z}\mathcal{T}$. More generally, $C_{4z}\mathcal{T}$ acts as an effective time-reversal symmetry, ensuring $\Omega_z(k_x,k_y)=-\Omega_z(-k_y,k_x)$ and resulting in a total Chern number of zero.

The spin Chern number $\mathcal{C}_s = (\mathcal{C}_{\uparrow} - \mathcal{C}_{\downarrow})/2$ \cite{2009prbE,2006prlD,2011prlY} is then evaluated by integrating $\Omega_z$. The analytical expression for $\mathcal{C}_{\uparrow}$ is:
\begin{equation}
\begin{aligned}
\mathcal{C}_{\uparrow}=-\mathcal{C}_{\downarrow} & = \frac{\mathrm{sgn}(t\lambda)}{2}\big[\mathrm{sgn}(u+2B)-\mathrm{sgn}(u-2B)\big], \\
& = \begin{cases} 0, & |u/B|>2, \\ \mathrm{sgn}(t\lambda B), & |u/B|<2, \end{cases}
\label{Chern}
\end{aligned}
\end{equation}
where $|u/B|=2$ marks the critical point at which $\mathcal{C}_s$ is not well-defined. The nonzero $\mathcal{C}_s$ depends solely on $u$ and $B$, and are the key parameters governing the topological phase transition. This result is consistent with the gap closing condition in Eq.~\eqref{eq:3}, validating the necessity of investigating the band inversion at X and Y points.

To confirm these findings, we consider two representative cases: $u/B = 2.20$ and $u/B = 1.60$. For $u/B = 2.20$, the system is in a trivial phase with no edge states, as shown in Fig.~\ref{fig:2}(b). For $u/B = 1.60$, spin-polarized helical edge states emerge, indicating a type-II QSH phase (Fig.~\ref{fig:2}(d)). Microscopically, $\Omega_z$ is nonzero around X/Y for both cases (Figs.~\ref{fig:2}(a) and (c)), but the compensated $\Omega_z$ around X/Y for $|u/B|>2$ leads to a trivial phase. The altermagnetic symmetry $C_{4z}\mathcal{T}$ ensures $\mathcal{C}_{\uparrow}=-\mathcal{C}_{\downarrow}$, and the resulting $\mathbb{Z}_2$ invariant in each spin species distinguishes this type-II QSH state from the conventional QSH effect in nonmagnetic systems.


Next, we investigate the effect of breaking the $C_{4z}\mathcal{T}$ symmetry. Unlike the topological phase transitions in nonmagnetic systems, where $\mathcal{T}$ is preserved under non-magnetic perturbation, altermagnets inherently break $\mathcal{T}$. Thus, breaking the crystalline symmetry component in $C_{4z}\mathcal{T}$ suffices to induce a transition, which can be achieved via external perturbations such as strain~\cite{2014prbS,2013japD}. 
To model this, we introduce two symmetry-breaking terms that contribute to sublattice inequivalence ($\mu_s$) and anisotropic SOC ($D'$). The modified Hamiltonian reads:
\begin{equation}
	\mathcal{H}' = \mu_s\tau_z\sigma_0 + D'(\cos k_x - \cos k_y)\tau_z\sigma_z.
	\label{eq:5}
\end{equation}
These terms explicitly break the $C_{4z}\mathcal{T}$ symmetry, lifting the degeneracy between valleys and enabling the gap to close and reopen at only a single valley.

As an illustrative example, we set $\mu_s = 0.6$ and $D' = 0$, and present the resulting band structure and edge states in Fig.~\ref{fig:2}(e) and (f). The Berry curvature $\Omega_z$ and edge state spectrum reveal that only the spin-up channel contributes to a nonzero Chern number, hosting a spin-polarized chiral edge mode. This phase, characterized by a sizable fundamental gap, represents a QAH state, demonstrating how topological properties can emerge from pristine altermagnets. In the following section, we will show the complete phase diagram in parameter space and how to realize these phases via strain.

\begin{figure}
	\centering 
	\includegraphics[width=0.5\textwidth]{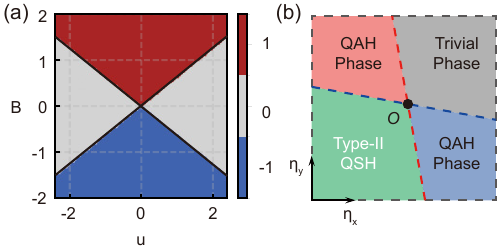}%
	\caption{\label{fig:3} (a) The phase diagram of the Chern number $\mathcal{C}_{\uparrow}$ for different values of u and B. It can be seen that the phase boundary corresponds to $|u/B|=2$, and the sign of $B$ determines the value of the $\mathcal{C}_{\uparrow}$. (b) Schematic phase diagram under various strain conditions, with the origin denoted by the point O. Red and blue lines indicate $|u_{\uparrow}/B_{\uparrow}|=2$ and $|u_{\downarrow}/B_{\downarrow}|=2$, respectively.}
\end{figure}

\textit{Topological phase diagram.}---Having established two distinct topological nontrivial phases, we now aim to construct the complete phase diagram based on our model. From Eq.~\eqref{eq:3} and the confirmed phases in Fig.~\ref{fig:2}, the critical condition is $|u/B| = 2$, with $\mathcal{C}_{\uparrow} = \mathcal{C}_{\downarrow} = 0$ for $|u/B| > 2$. By parameterizing the space of $u$ and $B$, we obtain the QSH phase diagram shown in Fig.~\ref{fig:3}(a). The two phase boundaries, defined by $|u/B| = 2$, divide the diagram into four regions. Besides the trivial phase, the regions with $\mathcal{C}_{\uparrow}$ of opposite signs correspond to band inversions occurring at different valleys (X or Y). Due to $C_{4z}\mathcal{T}$, the phase diagram for $\mathcal{C}_{\downarrow}$ is identical, with reversed signs.

When the $C_{4z}\mathcal{T}$ is broken, the gap values at X and Y become inequivalent, while the spin and valley are still locked in our discussion. Their expressions are modified to $\Delta_\mathrm{X} = 2|(u + \mu_s) - 2(B + D')|$ and $\Delta_\mathrm{Y} = 2|(u - \mu_s) - 2(B - D')|$. They can share the form in Eq.~\eqref{eq:3} with refined $u$ and $B$ as $u_{\uparrow/\downarrow} = u \pm \mu_s$ and $B_{\uparrow/\downarrow} = B \pm D'$. Treating $\mu_s$ and $D'$ as perturbations, the phase boundaries from the QSH phase diagram are still valid, and the system can host four distinct phases:

(i) Trivial Insulator: $|u_{\uparrow/\downarrow}/B_{\uparrow/\downarrow}| > 2$;

(ii) Type-II QSH Phase: $|u_{\uparrow/\downarrow}/B_{\uparrow/\downarrow}| < 2$, characterized by helical edge states with opposite spin polarizations;

(iii) QAH Phase I: $|u_{\uparrow}/B_{\uparrow}| < 2$ and $|u_{\downarrow}/B_{\downarrow}| > 2$, featuring a chiral edge state with spin-up polarization;

(iv) QAH Phase II: $|u_{\uparrow}/B_{\uparrow}| > 2$ and $|u_{\downarrow}/B_{\downarrow}| < 2$, featuring a chiral edge state with spin-down polarization.


\textit{Strain-induced topological phase transitions.}---With the established Hamiltonian, we choose strain to achive the four distinct phases. As shown analytically in Eq.~\eqref{Chern}, the Chern number for each spin channel depends solely on $u_{\uparrow/\downarrow}$ and $B_{\uparrow/\downarrow}$. To capture the strain effects, we express these parameters as functions of biaxial strain ($\eta_b$) and uniaxial strain ($\eta_u$)~\cite{2025prbT,2025aM}:
\begin{equation}
	\begin{aligned}
		u_{\uparrow/\downarrow}(\eta_b,\eta_u) &= u_0 + \delta_u\eta_b \pm \delta_{\mu_s}\eta_u, \\
		B_{\uparrow/\downarrow}(\eta_b,\eta_u) &= B_0 + \delta_B\eta_b \pm \delta_{D'}\eta_u,
	\end{aligned}
	\label{eq:4}
\end{equation}
where $\delta_i$ ($i = u, B, \mu_s, D'$) denote the strain coupling coefficients. We set $|u_0/B_0| = 2$ as the initial configuration at the critical point of the phase transition. To better characterize the lattice distortions, we relate $\eta_b$ and $\eta_u$ to uniaxial strains along $x$ and $y$ directions as $\eta_b = (\eta_x + \eta_y)/2$ and $\eta_u = (\eta_x - \eta_y)/2$.

The resulting phase diagram in the $\eta_x$-$\eta_y$ plane is shown in Fig.~\ref{fig:3}(b). It includes the topologically trivial phase (gray), the type-II QSH phase (green), and two QAH phases (red and blue). These regions are separated by two critical lines (red and blue) intersecting at the origin ($|u_0/B_0| = 2$). The slopes of the critical lines, $s_r$ and $s_b$, are the function of $\delta_i$ and given by:
\begin{equation}
s_r = \frac{1}{s_b} = -\frac{(\delta_u - 2\delta_B) + (\delta_{\mu_s} - 2\delta_{D'})}{(\delta_u - 2\delta_B) - (\delta_{\mu_s} - 2\delta_{D'})}.
\label{eq:k}
\end{equation}
The area of the QAH regions, proportional to $\delta_{\mu_s} - 2\delta_{D'}$, reflects the strain-induced symmetry breaking. This also provides a practical hint: in realistic systems, one should first apply biaxial strain to locate the origin, and then explore other phases around it. In the next section, we demonstrate a 2D altermagnet whose strain-driven phase diagram is precisely described by Fig.~\ref{fig:3}(b).

\begin{figure}
	\centering 
	\includegraphics[width=0.5\textwidth]{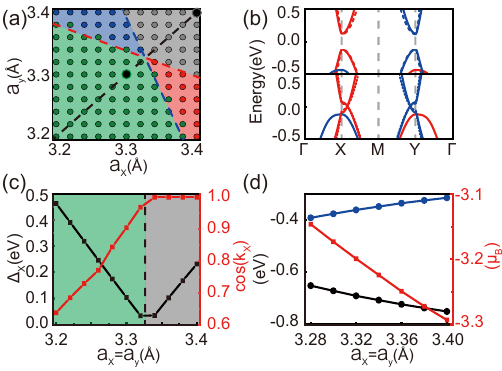}%
	\caption{\label{fig:4} (a) Phase diagram for different lattice constants. The black line denotes the diagonal, serving as an indicator of biaxial strain. (The data of blue and red line are in SM Part VI). (b) Band structure without SOC at a lattice constant of 3.40\AA and 3.30\AA (lower panel). Red and blue lines indicate spin-up and spin-down. Dashed lines correspond to the results of DFT calculations, whereas solid lines represent the model fits obtained from our theoretical framework (c) The black curve shows the energy difference at the X valley as a function of the lattice constant, while the red curve indicates the cosine value of the valley position. (d) The blue and black curves represent the fitted values of $u$ and $B$ at different lattice constants, respectively. The red curve shows the magnetic moment of a single magnetic atom as a function of lattice constants.}
\end{figure}

\textit{Material realization.}---Our model is based on a general square Lieb lattice, which serves as a versatile framework for describing a class of 2D altermagnets with $d$-wave Fermi surfaces. The topological phase transitions predicted in this work are expected to occur in materials exhibiting spin-polarized valleys at the X and Y points. First-principles studies have identified several candidates, including CrO~\cite{RN23,2023aplXC,2023ncmP}, Cr$_2$SO~\cite{2023aplS}, V$_2$Se$_2$O~\cite{2021ncH}, and Nb$_2$SeTeO~\cite{2025aF}. Among these, we select CrO as a representative example due to its simple elemental composition.

Using the pristine lattice parameters from Ref.~\cite{2023aplXC}, we calculate the band structure and topological properties of CrO, as shown in Fig.~\ref{fig:4}(b) lower panel. The spin-polarized valleys are located around the X and Y points, consistent with our model. With the relaxed lattice constant, band inversion occurs, confirming a type-II QSH phase. This is further validated by open boundary and Berry curvature calculations (see Supplemental Material~\cite{SM}). The Kondo-lattice model predicts an increase in $|u/B|$ with increasing lattice constant. We identify $a_x = a_y = 3.30 \ \text{\AA}$ as the starting point and apply a series of biaxial tensile strains to track the critical point. At approximately 0.6\% tensile strain ($a_x = a_y = 3.32 \ \text{\AA}$), the band gap closes, signaling a critical point. Then we choose $a_x = a_y = 3.32 \ \text{\AA}$ as the origin and conduct a series of calculations to obtain the phase diagram of CrO as presented in Fig.~\ref{fig:4}(a). Four phases are present under reasonable strain strengths, indicating that the topological phase transition is promising in experiments. The excellent agreement between the predicted phase diagram and first-principles results highlights the validity of our model in describing strain-induced topological phase transitions.

Focusing on the biaxial strain-induced transition, Fig.~\ref{fig:4}(c) shows that the phase transition is marked by the closing and reopening of the local gap at the X/Y points (black curve) and the deviation of the valleys from the X/Y points (red curve). The SOC opens a gap at the crossings near X/Y only when band inversion occurs, shifting the conduction and valence band extrema away from X/Y. Furthermore, our model provides insights into the strain dependence of $u$ and $B$. As shown in Fig.~\ref{fig:4}(d), the fitted $|u|$ decreases linearly with decreasing lattice constant, indicating a decrease in local magnetic moments. This trend is corroborated by the calculated magnetic moment of Cr atoms (red curve in Fig.~\ref{fig:4}(d)). In contrast, the fitted $|B|$ increases linearly with decreasing lattice constant. Together, $|u|$ and $|B|$ provide a reliable indicator of the topological phase transition, demonstrating again the predictive power of our model.

\textit{Conclusion.}---We have constructed a minimal model based on a 2D Lieb lattice to investigate the topological properties of altermagnets. At half-filling, we find that the spin Chern number $\mathcal{C}_s$ depends solely on the magnetic moment $u$ and the anisotropy strength $B$, with the critical condition $|u/B|=2$ marked by gap closure at the X and Y valleys. For $|u/B|<2$, the system hosts spin-polarized counterpropagating edge states, indicative of a type-II quantum spin Hall (QSH) phase. Breaking the $C_{4z}\mathcal{T}$ symmetry decouples the valleys, leading to chiral edge states with opposite spin polarizations and realizing two quantum anomalous Hall (QAH) phases with opposite Chern numbers. 

We further demonstrate that strain engineering, a non-magnetic tool, effectively tunes the model parameters to achieve these topological phases. First-principles calculations on monolayer CrO validate our predictions, showing four distinct phases within experimentally accessible strain regimes, in excellent agreement with the theoretical framework. Our findings highlight the potential of 2D altermagnets as a versatile platform for realizing and controlling topological phases, with strain providing an efficient and practical tuning mechanism.

\textit{Acknowledgments.}---This work was financially supported by the Key Project of the Natural Science Program of Xinjiang Uygur Autonomous Region (Grant No. 2023D01D03), the National Natural Science Foundation of China (Grant No. 52073308, No. 12304097 and No. 12164046), the Tianchi-Talent Project for Young Doctors of Xinjiang Uygur Autonomous Region (No. 51052300570) and the State Key Laboratory of Powder Metallurgy at Central South University. This work was carried out in part using computing resources at the High Performance Computing Center of Central South University.

\bibliography{letter1.bib}

\end{document}


\title{Supplemental Material for ``Multiple Topological Phases Controlled via Strain in Two-Dimensional Altermagnets"}

\author{Zesen Fu}
\thanks{These authors contributed equally to this work.}
\affiliation{School of Physics and Technology, Xinjiang University, Urumqi 830017, China}
\affiliation{School of Physics, Central South University, Changsha 410083, China}

\author{Mengli Hu}
\thanks{These authors contributed equally to this work.}
\affiliation{Leibniz Institute for Solid State and Materials Research, IFW Dresden, Helmholtzstraße 20, 01069 Dresden, Germany}

\author{Aolin Li}
\email{liaolin628@xju.edu.cn}
\affiliation{School of Physics and Technology, Xinjiang University, Urumqi 830017, China}

\author{Haiming Duan}
\affiliation{School of Physics and Technology, Xinjiang University, Urumqi 830017, China}

\author{Junwei Liu}
\affiliation{Department of Physics, The Hong Kong University of Science and Technology, Hong Kong, People’s Republic of China}

\author{Fangping Ouyang}
\email{ouyangfp06@tsinghua.org.cn}
\affiliation{School of Physics and Technology, Xinjiang University, Urumqi 830017, China}
\affiliation{School of Physics, Central South University, Changsha 410083, China}
\date{\today}

\maketitle

\tableofcontents

\renewcommand{\theequation}{S\arabic{equation}}
\renewcommand{\thefigure}{S\arabic{figure}}
\renewcommand{\thetable}{S\arabic{table}}
\setcounter{equation}{0}
\section{Hamiltonian}
The magnetic space group of the system can be generated by the following fundamental symmetry operations and their combinations:
\begin{equation*}
	E,M_x\mathcal{T},C_{4z}\mathcal{T},C_{2z}
\end{equation*}
Here, $E$ is the identity, $C_{2z}$ denotes twofold rotation, and the anti-unitary elements $\mathcal{T}M_x$ and $\mathcal{T}C_{4z}$ represent time-reversal symmetry combined with a mirror $M_x$ and fourfold rotation $C_{4z}$, respectively.
We choose four basis named: $\left|A,\uparrow\right>$, $\left|A,\downarrow\right>$, $\left|B,\uparrow\right>$ and $\left|B,\downarrow\right>$
Here, $A/B$ and $\uparrow/\downarrow$ denotes different sublattice and spin, respectively. Thus the symmetry in above can be expressed as:
\begin{equation*}
	\mathcal{T}=-i\sigma_y\tau_0\mathcal{K},
	M_x\mathcal{T}=\sigma_0\tau_0\mathcal{K},
	C_{4z}\mathcal{T}=\frac{i}{\sqrt{2}}(\sigma_y-\sigma_x)\tau_x\mathcal{K},
	C_{2z}=i\sigma_z\tau_0
\end{equation*}
In the above representations, $\sigma_{x,y,z}$ denotes the Pauli matrices acting on the spin degrees of freedom, while $\tau_{x,y,z}$ represents the Pauli matrices acting on the sublattice (orbital) space.
$\sigma_0$ and $\tau_0$ are both $2\times2$ identity operators. $\mathcal{K}$ is complex conjugation.

After including both nearest-neighbor and next-nearest-neighbor hoppings, we obtain the following set of linearly independent matrices under the symmetry constraints:

\begin{equation*}
	L_1=\left(\begin{matrix}1&0&0&0\\0&0&0&0\\0&0&0&0\\0&0&0&1\\\end{matrix}\right)
	L_2=\left(\begin{matrix}0&0&0&0\\0&1&0&0\\0&0&1&0\\0&0&0&0\\\end{matrix}\right)
\end{equation*}

\begin{equation*}
	L_3=\left(\begin{matrix}\cos{k_x}&0&0&0\\0&0&0&0\\0&0&0&0\\0&0&0&\cos{k_x}\\\end{matrix}\right)
	L_4=\left(\begin{matrix}0&0&0&0\\0&\cos{k_x}&0&0\\0&0&\cos{k_x}&0\\0&0&0&0\\\end{matrix}\right)
\end{equation*}

\begin{equation*}
	L_5=\left(\begin{matrix}\cos{k_y}&0&0&0\\0&0&0&0\\0&0&0&0\\0&0&0&\cos{k_y}\\\end{matrix}\right)
	L_6=\left(\begin{matrix}0&0&0&0\\0&\cos{k_y}&0&0\\0&0&\cos{k_y}&0\\0&0&0&0\\\end{matrix}\right)
\end{equation*}

\begin{equation*}
	L_7=\cos{\frac{k_x}{2}}\cos{\frac{k_y}{2}}\sigma_0\tau_x=\left(\begin{matrix}0&0&\cos{\frac{k_x}{2}}\cos{\frac{k_y}{2}}&0\\0&0&0&\cos{\frac{k_x}{2}}\cos{\frac{k_y}{2}}\\\cos{\frac{k_x}{2}}\cos{\frac{k_y}{2}}&0&0&0\\0&\cos{\frac{k_x}{2}}\cos{\frac{k_y}{2}}&0&0\\\end{matrix}\right)	
\end{equation*}

\begin{equation*}
	L_8=\sin{\frac{k_x}{2}}\sin{\frac{k_y}{2}}\sigma_z\tau_y=\left(\begin{matrix}0&0&-i\sin{\frac{k_x}{2}}\sin{\frac{k_y}{2}}&0\\0&0&0&-i\sin{\frac{k_x}{2}}\sin{\frac{k_y}{2}}\\-i\sin{\frac{k_x}{2}}\sin{\frac{k_y}{2}}&0&0&0\\0&-i\sin{\frac{k_x}{2}}\sin{\frac{k_y}{2}}&0&0\\\end{matrix}\right)	
\end{equation*}

The symmetry constraints can be written as:
\begin{equation*}
	U_gL\left(\bm{k}\right)U_g^{-1}=L(R\cdot\bm{k})
\end{equation*}
Here, $g$ denotes an element of the symmetry group, $U_g$ is its matrix representation, and $R$ represents the corresponding transformation induced by $g$ in reciprocal space. The Hamiltonian can then be written as:
\begin{equation}
	\mathcal{H}=\sum_{i}^{8}{l_iL_i}
	\label{eq:S0}
\end{equation}
Here, $l_i$ denote the coefficients in the linear combination. To make the structure of the Hamiltonian more compact and to clarify the physical meaning of each parameter, we reorganize the basis of the linearly independent matrices and redefine the combination scheme. This yields the final form of the Hamiltonian:
\begin{equation}
	\begin{aligned}
		\mathcal{H}&(k_x,k_y)= 
		\Bigl[\mu + A(\cos k_x + \cos k_y)\Bigr]\tau_0\sigma_0 + 
		B\Bigl[\cos k_x - \cos k_y\Bigr]\tau_z\sigma_0 + t\cos \frac{k_x}{2}\cos \frac{k_y}{2}\tau_x\sigma_0\\
		& + \lambda\sin \frac{k_x}{2}\sin \frac{k_y}{2}\tau_y\sigma_z + C\Bigl[\cos k_x - \cos k_y\Bigr]\tau_0\sigma_z+ \Bigl[u + D(\cos k_x + \cos k_y)\Bigr]\tau_z\sigma_z
	\end{aligned}
	\label{eq:S1}
\end{equation} 
\begin{equation}
\mathcal{H}\left(k_x,k_y\right)=H_1\tau_0\sigma_0+H_2\tau_z\sigma_0+H_3\tau_x\sigma_0+H_4\tau_y\sigma_z+H_5\tau_0\sigma_z+H_6\tau_z\sigma_z
	\label{eq:S2}
\end{equation} 

By comparing Eq~\eqref{eq:S1} and Eq~\eqref{eq:S2}, we can derive the terms $H_1$ to $H_6$.
Here we provide a detailed explanation of the physical meanings of the parameters in the model. The roles of $A$ and $B$ (isotropic and anisotropic hoppings of NNNs), $t$ (NN hopping), and $\lambda$ (SOC) have already been discussed in the main text. We now focus on clarifying the physical significance of $C$ and $D$.

In the main text, the spin-dependent part of the NNN hopping involves the parameters $C$ and $D$. As shown in Eq.~\eqref{eq:S1}, both are associated with spin-dependent terms. If $C$ and $D$ were solely due to SOC, then in the absence of SOC we would have $C=D=\lambda=0$, and the Hamiltonian would reduce to a form similar to that in Refs.~\cite{RN45,2025aF}, where the system exhibits particle–hole symmetry.
This can be seen directly from Eqs.~\eqref{Eup} and \eqref{Edn}: when $C=D=0$, $H_5=0$, and for every valence-band energy $E^{v}_{\uparrow/\downarrow}(k_x,k_y)$ there exists a corresponding conduction-band energy $E^{c}_{\uparrow/\downarrow}(k_x,k_y)=-E^{v}_{\uparrow/\downarrow}(k_x,k_y)$.
However, in certain altermagnets such symmetry is generally absent. Thus, $C$ and $D$ cannot be fully attributed to SOC. Instead, part of their contribution arises from spin-dependent differences in electronic hopping mediated by local magnetic atoms. This motivates our description in terms of a Kondo-lattice model. Consequently, even without SOC, we still have $C,D \neq 0$ in our discussion.

In Eq.~\eqref{eq:S2}, we can find $\left[\mathcal{H}\left(k_x,k_y\right),\sigma_z\right]=0$, which means spin is conserved. Thus, we can analytically solve for the energy eigenvalues corresponding to each band by diagonalizing the Hamiltonian for each spin.
Here, we show the result:
\begin{equation}
	E_{\uparrow}=H_1+H_5\pm\sqrt{H_3^2+H_4^2+\left(H_2+H_6\right)^2}
    \label{Eup}
\end{equation}
\begin{equation}
	E_{\downarrow}=H_1-H_5\pm\sqrt{H_3^2+H_4^2+\left(H_2-H_6\right)^2}
    \label{Edn}
\end{equation}
Due to the symmetry $C_{4z}\mathcal{T}$, $E_{\uparrow}\left(k_x,k_y\right)=E_{\downarrow}(-k_y,k_x)$, for the analysis of valley and band properties, we can focus solely on the spin-up sector of the Hamiltonian.

The energy separation for spin up is $\Delta=2\sqrt{H_3^2+H_4^2+\left(H_2+H_6\right)^2}$.
The function $\partial\Delta/\partial k_{x/y}=0$ identifies the possible positions of valleys in the Brillouin zone.
The solutions are $\Gamma\left(0,0\right)$,$\mathrm{X}\left(\pi,0\right)$,$\mathrm{Y}\left(0,\pi\right)$ and $\mathrm{M}(\pi,\pi)$.
In this case, we assume the effective SOC term $\lambda=0$. The energy separation at the four valley points are:
\begin{align*}
	\Delta_\Gamma=2\sqrt{t^2+\left(u+2D\right)^2},
	\Delta_\mathrm{X}=2|u-2B|,
	\Delta_\mathrm{M}=2|u-2D|,
	\Delta_\mathrm{Y}=2|u+2B|.
\end{align*}
Since the nearest-neighbor hopping parameter t is typically large, the valleys generally do not appear at the $\Gamma$ point.
Due to symmetry, band degeneracy typically occurs at the M point.
In this case, we exclude it from the valley analysis and focus only on valleys that may form at X and Y.
To locate the dominant valley, we compare the energy separation at X and Y:
\begin{align*}
	\begin{cases}
		\mathrm{if}\ u/B>0, \Delta_\mathrm{X}<\Delta_\mathrm{Y}\\
		\mathrm{if}\ u/B<0, \Delta_\mathrm{X}>\Delta_\mathrm{Y}
	\end{cases}
\end{align*}
Thus, we give an example to solve the condition that spin-up valley located at X. 
Along the high-symmetry line XM, the energy separation for spin up can be expressed as
\begin{align*}
	\Delta_\mathrm{XM}=2\left|u-D-B+\left(D-B\right)\cos{k_y}\right|,\cos{k_y}\in\left[-1,1\right],
\end{align*}
and similarly, along the YM line:
\begin{align*}
	\Delta_\mathrm{YM}=2\left|u-D+B+\left(D+B\right)\cos{k_x}\right|,\cos{k_x}\in\left[-1,1\right]
\end{align*}
Thus, the constraints for spin-up valley located at X are:
\[
\left\{
\begin{array}{rl}
	(u-2B)(D-B)   &< 0 \\
	(u-2B)(u-2D)  &> 0 \\
	(u+2B)(u-2D)  &> 0 \\
	\multicolumn{2}{c}{u/B > 0}
\end{array}
\right.
\]
The solution is:
\begin{equation}
	\begin{cases}
		u/B>2\\
		D/B<1
	\end{cases}
	\label{eq:S5}
\end{equation}
Thus, under the condition above, the spin-up valley is locking at X, and the energy separation at X for spin-up ($\Delta_\mathrm{X}=|u-2B|$) can represent the energy gap for this system.
Similarly, the solution for the spin-up valley locked at the Y point is:
\begin{equation}
	\begin{cases}
		u/B<-2\\
		D/B>-1
	\end{cases}
	\label{eq:S6}
\end{equation}

\begin{figure}
	\centering 
	\includegraphics[width=0.8\textwidth]{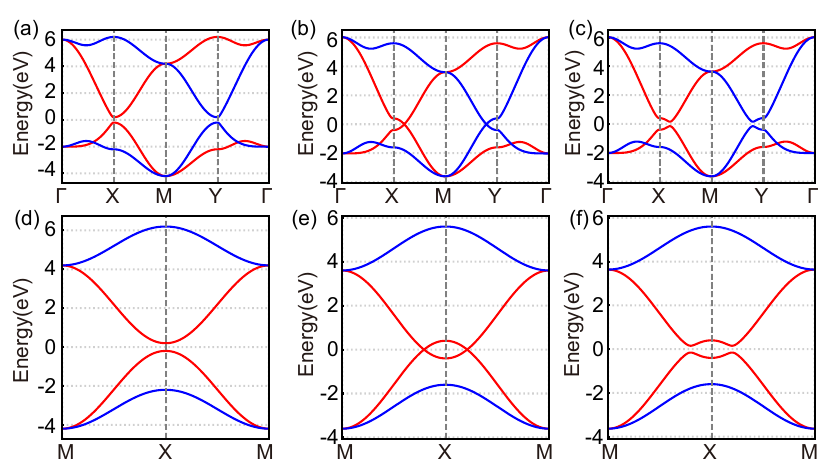}%
	\caption{\label{fig:S1} 
		The model parameters are set as follows:$A=0.5$, $B=-1$, $C=0.5$, $D=1$, $\mu=1$ and $t=4$. (a) and (b) are the band structures of $u=-2.2$ and $u=-1.6$ with $\lambda=0$, respectively. (c) is the band structure of $u=-1.6$ and $\lambda=0.5$. The band structures long the M–X–M path for them are shown in (d), (e) and (f). The red and blue lines represent spin-up and spin-down, respectively.}
\end{figure}

Taking the spin-up valley locate at X as an example, when the ratio u/B decreases from greater than 2 to less than 2, the energy separation at the X point continuously shrinks. It closes exactly at $u/B=2$, and then reopens. Due to the $C_{4z}\mathcal{T}$ symmetry, a similar transition will happen in the Y (spin-down) valley.
During this process, the valley position shifts away from X. This new valley position is determined by solving the condition: $\Delta_\mathrm{XM}=0$
\begin{align*}
	\left(k_x,k_y\right)=\left(\pi,\pm a c r c o s\left(\frac{B+D-u}{D-B}\right)\right)
\end{align*}
Figs.~\ref{fig:S1}(a) and (b) illustrate the band structure evolution during this process. Figs.~\ref{fig:S1}(d) and (e) present the band structures traced along the high-symmetry path MXM. They clearly show the transition from a single valley at the high-symmetry point X to two split valleys located away from X and the system's transition from an insulating phase to a semimetal. Figs.~\ref{fig:S1}(c) and (f) show the band structure after including the effective SOC term $\lambda$. It opens a gap at the degenerate point.

\section{Spin Chern number}
The closing and reopening of the bandgap, often signal the emergence of nontrivial topological edge states. To investigate the topological nature of the system, we analyze the Berry curvature, which captures the geometric phase structure of the Bloch bands.
The Berry curvature for the nth band is given by:
\begin{equation}
	\Omega_n\left(\bm{k}\right)=i\sum_{m\neq n}\frac{\left\langle u_n\left(\bm{k}\right)\right|\partial_{k_x}H\left(\bm{k}\right)\left|u_m\left(\bm{k}\right)\right\rangle\left\langle u_m\left(\bm{k}\right)\right|\partial_{k_y}H\left(\bm{k}\right)\left|u_n\left(\bm{k}\right)\right\rangle-\left(k_x\longleftrightarrow k_y\right)}{\left(E_m\left(\bm{k}\right)-E_n\left(\bm{k}\right)\right)^2}
    \label{berry4}
\end{equation}
Here, $\left|u_n\left(\bm{k}\right)\right\rangle$ is the periodic part of the Bloch wavefunction for band $n$, $H\left(\bm{k}\right)$ is the Bloch Hamiltonian, $E_n\left(\bm{k}\right)$ is the eigenenergy of band $n$.
The Chern number of the $n$th band is 
\begin{equation}
	\mathcal{C}_n=\frac{1}{2\pi}\int_{BZ}{\Omega_n\left(\bm{k}\right)d^2k}.
	\label{eq:chern}
\end{equation}
Due to the presence of $\mathcal{T}C_{4z}$ symmetry, $\Omega\left(k_x,k_y\right)=-\Omega\left(-k_y,k_x\right)$, the total Chern number of the system is strictly constrained to be zero.
At the same time, to capture the topological nature of this phase with time-reversal symmetry ($\mathcal{T}$) borken, we introduce the concept of the spin Chern number.

For a system that breaks the time-reversal symmetry but retains spin conservation, i.e.,$\left[\mathcal{H},S_\alpha\right]=0$.
Here, $\mathcal{H}$ is the Hamiltonian of the system and $S_\alpha$ is the spin operator in the $\alpha$-direction.
The commutation between the Hamiltonian and the spin operator allows the Hamiltonian to be discretized into two separate parts based on the spin being positive or negative in the $\alpha$-direction, namely $\mathcal{H}_+$ and $\mathcal{H}_-$.
For each of these independent Hamiltonians, the corresponding Berry curvature and Chern number can be calculated.
The total Chern number can then be divided into two parts: $\mathcal{C}_+$ and $\mathcal{C}_-$. 
By calculating $\mathcal{C}_s={(\mathcal{C}}_+-\mathcal{C}_-)/2$, one can determine whether the QSHE is present.

\begin{figure}
	\centering 
	\includegraphics[width=0.6\textwidth]{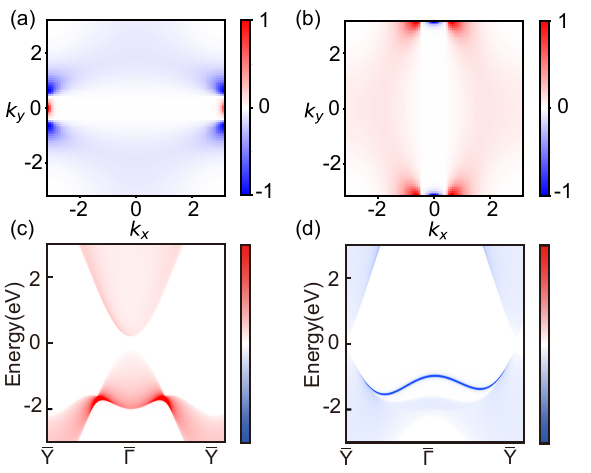}%
	\caption{\label{fig:S3} 
		(a) Berry curvature and (c) edge state for spin-up at $u=-2.2$. (b) Berry curvature and (d) edge state for spin-down at $u=-2.2$. In plotting the Berry curvature, we performed a normalization such that $\Omega_z$ is divided by $\mathrm{Max}(\Omega_z)$ for positive values ($\Omega_z > 0$) and by $\mathrm{Min}(\Omega_z)$ for negative values ($\Omega_z < 0$). Moreover, Spin-polarized LDOS (edge states) is obtained by subtracting the spin-down component from the spin-up component: deeper red indicates a larger contribution from spin-up states, while deeper blue corresponds to spin-down states.}
\end{figure}

Reviewing the system Hamiltonian given by Eq.~\eqref{eq:S2}, we notice that $\left[\mathcal{H},\sigma_z\right]=0$, which means we can decompose the Hamiltonian into two independent parts, $H_{\uparrow}$ and $H_{\downarrow}$, corresponding to spin-up and spin-down states, respectively.
\begin{equation}
	H_{\uparrow}=\left(H_1+H_5\right)\tau_0+H_3\tau_x+H_4\tau_y+\left(H_2+H_6\right)\tau_z
	\label{eq:S7}
\end{equation}
\begin{equation}
	H_{\downarrow}=\left(H_1-H_5\right)\tau_0+H_3\tau_x-H_4\tau_y+\left(H_2-H_6\right)\tau_z
\end{equation}
For $u/B>2$, Figs.~\ref{fig:S3}(a) and (c) display the Berry curvature and edge state spectrum for the spin-up sector.
The Berry curvature integrates to zero, and no edge states are present.
Figs.~\ref{fig:S3}(b) and (d) show the corresponding results for the spin-down sector, also exhibiting trivial topology.

\begin{figure}
	\centering 
	\includegraphics[width=0.6\textwidth]{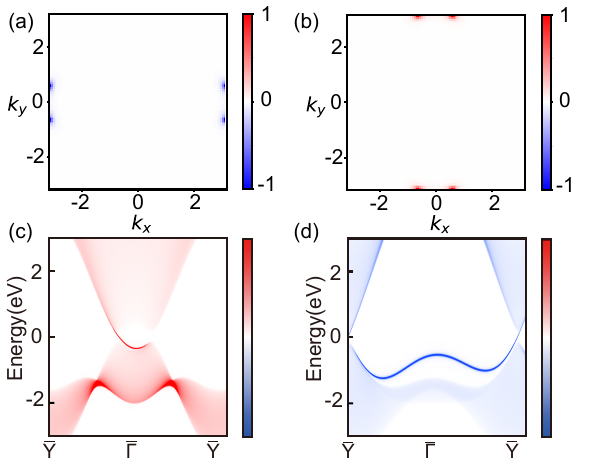}%
	\caption{\label{fig:S4} 
		(a) Berry curvature and (c) edge state for spin-up at $u=-1.6$. (b) Berry curvature and (d) edge state for spin-down at $u=-1.6$. In plotting the Berry curvature, we performed a normalization such that $\Omega_z$ is divided by $\mathrm{Max}(\Omega_z)$ for positive values ($\Omega_z > 0$) and by $\mathrm{Min}(\Omega_z)$ for negative values ($\Omega_z < 0$). Moreover, Spin-polarized LDOS (edge states) is obtained by subtracting the spin-down component from the spin-up component: deeper red indicates a larger contribution from spin-up states, while deeper blue corresponds to spin-down states.}
\end{figure}

In contrast, when $u/B<2$, Figs.~\ref{fig:S4}(a) and (c) show the Berry curvature and edge states for the spin-up sector. A Chern number of $\mathcal{C}_{\uparrow}=-1$ is obtained, and clearly resolved chiral edge states emerge. Figs.~\ref{fig:S4}(b) and (d) illustrate the spin-down sector, with a Chern number $\mathcal{C}_{\downarrow}=+1$ and similarly visible edge states. The resulting spin Chern number is $\mathcal{C}_s=-1$, which indicates the system has entered a quantum spin Hall phase. The coexistence of counter-propagating spin-polarized edge states in opposite spin channels, as observed in the edge spectra, provides direct evidence for the presence of this topological state.

\section{Phase Change Conditions}
As shown in Eq.~\eqref{eq:S7}, the spin-up Hamiltonian can be written in the form:
\begin{align*}
	H_{\uparrow}=H_0+\mathbf{d}\left(\bm{k}\right)\cdot\bm{\tau}
\end{align*}
Here, $H_0=H_1+H_5$ is a scalar to describe energy shift, and $\mathbf{d}\left(\bm{k}\right)=\left(H_3,H_4,H_2+H_6\right)$ is a momentum-dependent vector.
Since the topological properties of the system---such as Berry curvature and Chern number---are determined solely by the structure of $\mathbf{d}\left(\bm{k}\right)$, we can safely ignore $H_0$ in our topological analysis.

For a two band Hamiltonian, the Berry curvature \eqref{berry4}(for lower energy) can be expressed as:
\begin{equation}
	\Omega_z(k_x,k_y)=-2\mathbf{Im}\frac{\left<0\right|\partial_{k_x}H(\bm{k})\left|1\right>\left<1\right|\partial_{k_y}H(\bm{k})\left|0\right>}{4\mathbf{d}^2}
	\label{berry}
\end{equation}
Here, $\left|0\right>$ and $\left|1\right>$ represent the occupied state and the unoccupied state. In addition, we can explicitly compute the analytical expression of the Berry curvature.
\begin{equation}
	\Omega_{z}^{\uparrow}(k_x,k_y)=\frac{t\lambda}{16}\frac{B(\sin{k_x}^2+\sin{k_y}^2-2\cos{k_x}\cos{k_y}+2)+D(\sin{k_x}^2-\sin{k_y}^2)+u(\cos{k_x}-\cos{k_y})}{|(t\cos{\frac{k_x}{2}}\cos{\frac{k_y}{2}})^2+(\lambda\sin{\frac{k_x}{2}}\sin{\frac{k_y}{2}})^2+(u+B(\cos{k_x}-\cos{k_y})+D(\cos{k_x}+\cos{k_y}))^2|^{\frac{3}{2}}}
	\label{berry3}
\end{equation}
By integrating the Berry curvature from Eq.~\eqref{eq:chern}, we can obtain the $\mathcal{C}_{\uparrow}$ for $H_\uparrow$:
\begin{equation}
\mathcal{C}_{\uparrow}=\frac{\mathrm{sgn}(t\lambda)}{2}\big[\mathrm{sgn}(u+2B)-\mathrm{sgn}(u-2B)\big]
\label{Chern}
\end{equation}
Thus, we can get the phase change condition: $|u/B|=2$.
$|u/B|>2$ denotes a trivial state, with $\mathcal{C}_{\uparrow}=-\mathcal{C}_{\downarrow}=0$, while $|u/B|>2$ denotes a QSH state with $\mathcal{C}_{+}=-\mathcal{C}_{-}=\mathrm{sgn}(tB\lambda)$.
At the same time, we also find that the phase change condition exactly corresponds to the condition where the bands gap close at the X or Y valley. Therefore, it is precisely the closing and reopening of the bands that leads to band inversion, thereby causing the topological phase transition.

The above analysis has been carried out under the assumption that the $C_{4z}\mathcal{T}$ symmetry is preserved. We now turn to the case where this symmetry is broken.
Using symmetry analysis under the reduced symmetry group, we rewrite the Hamiltonian as a linear combination of allowed terms (similar the Eq.~\eqref{eq:S0}):
\begin{align*}
	{l_1L}_1\rightarrow l_{1,\uparrow}\left(\begin{matrix}1&0&0&0\\0&0&0&0\\0&0&0&0\\0&0&0&0\\\end{matrix}\right)+l_{1,\downarrow}\left(\begin{matrix}0&0&0&0\\0&0&0&0\\0&0&0&0\\0&0&0&1\\\end{matrix}\right)
\end{align*}
\begin{align*}
	{l_2L}_2\rightarrow l_{2,\uparrow}\left(\begin{matrix}0&0&0&0\\0&0&0&0\\0&0&1&0\\0&0&0&0\\\end{matrix}\right)+l_{2,\downarrow}\left(\begin{matrix}0&0&0&0\\0&1&0&0\\0&0&0&0\\0&0&0&0\\\end{matrix}\right)
\end{align*}
\begin{align*}
	{l_3L}_3\rightarrow l_{3,\uparrow}\left(\begin{matrix}\cos{k_x}&0&0&0\\0&0&0&0\\0&0&0&0\\0&0&0&0\\\end{matrix}\right)+l_{3,\downarrow}\left(\begin{matrix}0&0&0&0\\0&0&0&0\\0&0&0&0\\0&0&0&\cos{k_x}\\\end{matrix}\right)
\end{align*}
\begin{align*}
	l_4L_4\rightarrow l_{4,\uparrow}\left(\begin{matrix}0&0&0&0\\0&0&0&0\\0&0&\cos{k_x}&0\\0&0&0&0\\\end{matrix}\right)+l_{4,\downarrow}\left(\begin{matrix}0&0&0&0\\0&\cos{k_x}&0&0\\0&0&0&0\\0&0&0&0\\\end{matrix}\right)
\end{align*}
\begin{align*}
	{l_5L}_5\rightarrow l_{5,\uparrow}\left(\begin{matrix}\cos{k_y}&0&0&0\\0&0&0&0\\0&0&0&0\\0&0&0&0\\\end{matrix}\right)+l_{5,\downarrow}\left(\begin{matrix}0&0&0&0\\0&0&0&0\\0&0&0&0\\0&0&0&\cos{k_y}\\\end{matrix}\right)
\end{align*}
\begin{align*}
	l_6L_6\rightarrow l_{6,\uparrow}\left(\begin{matrix}0&0&0&0\\0&0&0&0\\0&0&\cos{k_y}&0\\0&0&0&0\\\end{matrix}\right)+l_{6,\downarrow}\left(\begin{matrix}0&0&0&0\\0&\cos{k_y}&0&0\\0&0&0&0\\0&0&0&0\\\end{matrix}\right)
\end{align*}
\begin{align*}
	tL_7=t\cos{\frac{k_x}{2}}\cos{\frac{k_y}{2}}\sigma_0\tau_x\rightarrow t_{\uparrow}\cos{\frac{k_x}{2}}\cos{\frac{k_y}{2}}\frac{1+\sigma_z}{2}\tau_x+t_{\downarrow}\cos{\frac{k_x}{2}}\cos{\frac{k_y}{2}}\frac{1-\sigma_z}{2}\tau_x
\end{align*}
\begin{align*}
	{\lambda L}_8=\lambda\sin{\frac{k_x}{2}}\sin{\frac{k_y}{2}}\sigma_z\tau_y\rightarrow\lambda_{\uparrow}\sin{\frac{k_x}{2}}\sin{\frac{k_y}{2}}\frac{1+\sigma_z}{2}\tau_y-\lambda_{\downarrow}\sin{\frac{k_x}{2}}\sin{\frac{k_y}{2}}\frac{1-\sigma_z}{2}\tau_y
\end{align*}
\begin{align*}
	H=\sum_{i}^{8}{l_iL_i}\rightarrow H=\sum_{i}^{8}{l_{i,\uparrow}L_{i,\uparrow}}+\sum_{i}^{8}{l_{i,\downarrow}L_{i,\downarrow}}
\end{align*}
Under such conditions, the Hamiltonian, originally expressed as a linear combination of 8 matrices, becomes a linear combination of 16 matrices.
However, we find that spin conservation is still preserved after breaking the $C_{4z}\mathcal{T}$ symmetry. Therefore, the full Hamiltonian is divided into two separate parts. The difference now is that these two parts no longer share the same set of parameters, but instead possess their own distinct parameter sets.
\begin{equation}
	\begin{aligned}
		H_{\uparrow}=& 
		\Bigl[\mu_{\uparrow} + A_{\uparrow}(\cos k_x + \cos k_y)\Bigr]\tau_0 + 
		B_{\uparrow}\Bigl[\cos k_x - \cos k_y\Bigr]\tau_z + t_{\uparrow}\cos \frac{k_x}{2}\cos \frac{k_y}{2}\tau_x\\
		& + \lambda_{\uparrow}\sin \frac{k_x}{2}\sin \frac{k_y}{2}\tau_y + C_{\uparrow}\Bigl[\cos k_x - \cos k_y\Bigr]\tau_0+ \Bigl[u_{\uparrow} + D_{\uparrow}(\cos k_x + \cos k_y)\Bigr]\tau_z
	\end{aligned}
	\label{eq:S10}
\end{equation} 
\begin{equation}
	\begin{aligned}
		H_{\downarrow}=& 
		\Bigl[\mu_{\downarrow} + A_{\downarrow}(\cos k_x + \cos k_y)\Bigr]\tau_0 + 
		B_{\downarrow}\Bigl[\cos k_x - \cos k_y\Bigr]\tau_z + t_{\downarrow}\cos \frac{k_x}{2}\cos \frac{k_y}{2}\tau_x\\
		& - \lambda_{\downarrow}\sin \frac{k_x}{2}\sin \frac{k_y}{2}\tau_y - C_{\downarrow}\Bigl[\cos k_x - \cos k_y\Bigr]\tau_0- \Bigl[u_{\downarrow} + D_{\downarrow}(\cos k_x + \cos k_y)\Bigr]\tau_z
	\end{aligned}
	\label{eq:S11}
\end{equation} 
\begin{equation}
	H=H_{\uparrow}(1+\sigma_z)/2+H_{\downarrow}(1-\sigma_z)/2
\end{equation}
To compare with the Hamiltonian in Eq.~\eqref{eq:S1}, and only consider the terms correspond to the topological transition, we get:
\begin{equation*}
	H=\mathcal{H}+H'
\end{equation*}
\begin{equation}
H'=\mu_s\tau_z\sigma_0+D'(\cos k_x-\cos k_y)\tau_z\sigma_z.
	\label{eq:S14}
\end{equation}
Here, $\mathcal{H}$ denotes the Hamiltonian that preserves the $\mathcal{T}C_{4z}$ symmetry, $H$ is the Hamiltonian after the symmetry is broken, and $H'$ represents the additional terms induced by the symmetry breaking with sublattice inequivalence: $\mu_s$, and anisotropic SOC: $D'$. Thus, compare Eq.\eqref{eq:S10} and Eq.\eqref{eq:S11} with Eq.\eqref{eq:S1} and Eq.\eqref{eq:S14} we get:
\begin{align*}
	u_{\uparrow/\downarrow}=u\pm\mu_s,B_{\uparrow/\downarrow}=B\pm D'
\end{align*}
The gap for the spin-up or spin-down valley, which was originally given by $|u-2B|$ in the presence of unbroken symmetry, is now modified to $|u_{\uparrow}-2B_{\uparrow}|$ for X (spin-up) valley, and $|u_{\downarrow}-2B_{\downarrow}|$ for Y (spin-down) valley.
As demonstrated in the previous proof, the conditions for band gap closing are analogous to those for a topological phase transition. Thus, the criteria for the phase transition are modified as follows:
\begin{align*}
	\left|u/B\right|=2\rightarrow\left|u_{\uparrow}/B_{\uparrow}\right|=2\ \mathrm{and}\ \left|u_{\downarrow}/B_{\downarrow}\right|=2.
\end{align*}
In this case, we can get 4 distinct phases:

(a)Trivial Insulator:  $|u_{\uparrow}/B_{\uparrow}| > 2$ and $|u_{\downarrow}/B_{\downarrow}| > 2$; 

(b)QAH Phase I: $|u_{\uparrow}/B_{\uparrow}| > 2$ and $|u_{\downarrow}/B_{\downarrow}| < 2$, yielding $\mathcal{C}_{\uparrow} = 0$ and $\mathcal{C}_{\downarrow} = -\mathrm{sgn}(tB\lambda)$, so the total Chern number is $\mathcal{C} = -\mathrm{sgn}(tB\lambda)$; 

(c)QAH Phase II: $|u_{\uparrow}/B_{\uparrow}| < 2$ and $|u_{\downarrow}/B_{\downarrow}| > 2$, yielding $\mathcal{C}_{\uparrow} = \mathrm{sgn}(tB\lambda)$ and $\mathcal{C}_{\downarrow} = 0$, so the total Chern number is $\mathcal{C} = \mathrm{sgn}(tB\lambda)$, opposite to Phase I; 

(d)Type-II QSH Phase: $|u_{\uparrow}/B_{\uparrow}| < 2$ and $|u_{\downarrow}/B_{\downarrow}| < 2$, yielding two different spin-polarized edge states at different k-points.

\section{Strain Perturbation Matrix}
We now examine the effect of strain on the full Hamiltonian.
We restrict the applied strain to be along the x- and y-directions, denoted by $\eta_x$ and $\eta_y$, respectively. Under such strain, the lattice constants are modified as:$a_x=a_0\left(1+\eta_x\right)$, $a_y=a_0\left(1+\eta_y\right)$.

Firstly, we consider the case of biaxial strain ($\eta_x=\eta_y=\eta$). Since biaxial strain does not break the original symmetry and we only focus on how the strain induces the phase transition, the parameters irrelevant to the phase transition are neglected. Thus, the Hamiltonian after strained can be written as:
\begin{equation*}
	\mathcal{H}(\eta)=\mathcal{H}_0+\mathcal{H}'
\end{equation*}
\begin{equation}
	\mathcal{H}'(\eta)=\Delta u(\eta)\tau_z\sigma_z+ \Delta B(\eta)(\cos{k_x}-\cos{k_y})\tau_z\sigma_0
	\label{eq:S15}
\end{equation}
Here, $\mathcal{H}$ and $\mathcal{H}_0$ denote the Hamiltonian after and before the application of the biaxial strain, respectively.
$\Delta u(\eta)$ and $\Delta B(\eta)$ are both functions of $\eta$, describing the changes in the parameters $u$ and $B$ induced by the biaxial strain.
Since the applied strain is small, we can expand the parameters to first order, yielding $\Delta u(\eta)=\frac{\partial u}{\partial\eta} \eta=\delta_u\eta$, $\Delta B(\eta)=\frac{\partial B}{\partial\eta} \eta=\delta_B\eta$ where $\delta_u=\frac{\partial u}{\partial\eta}$ and $\delta_B=\frac{\partial B}{\partial\eta}$ denote the linear response coefficient of $u$ and $B$ to strain, respectively.

Similarly, we also consider a different type of anti-symmetric strain with $\eta_x=-\eta_y=\eta$.
Compared with Eq.\eqref{eq:S15}, this strain introduces an additional term that breaks the altermagnetic symmetry, similar to Eq.\eqref{eq:S14}. Taking into account the inherent relation under this anti-symmetric operation,
\begin{align*}
	C_{4z}H'\left(\eta,-\eta\right)C_{4z}^{-1}+H'\left(\eta,-\eta\right)=0.
\end{align*}
Here, $H'$ denotes the additional term introduced by the anti-symmetric strain. Then, we can write the form of the Hamiltonian under such strain as:
\begin{equation*}
	H(\eta)=\mathcal{H}_0+H'
\end{equation*}
\begin{equation}
	H'(\eta)=\mu_s(\eta)\tau_z\sigma_0+ D'(\eta)(\cos{k_x}-\cos{k_y})\tau_z\sigma_z
	\label{eq:S16}
\end{equation}
Here, $H$ and $\mathcal{H}_0$ denote the Hamiltonian after and before the application of the anti-symmetric strain, respectively.
$\mu_s(\eta)$ and $D'(\eta)$ are both functions of $\eta$, describing symmetry broken term (Eq.~\eqref{eq:S14}) induced by the anti-symmetric strain.
Since the applied strain is small, we can expand the parameters to first order, yielding $ \mu_s(\eta)=\frac{\partial\mu_s}{\partial\eta} \eta=\delta_{\mu_s}\eta$, $D'(\eta)=\frac{\partial D'}{\partial\eta} \eta=\delta_{D'}\eta$ where $\delta_{\mu_s}=\frac{\partial \mu_s}{\partial\eta}$ and $\delta_{D'}=\frac{\partial D'}{\partial\eta}$ characterize the rates of change induced by the additional symmetry-breaking terms associated with the anti-symmetric strain.

For a general uniaxial strain ($\eta_x,\eta_y$), the associated perturbation matrix can be systematically decomposed into two parts: a symmetric (biaxial) contribution ($\eta_b$) and an anti-symmetric ($\eta_u$) contribution.
\begin{equation}
	H'(\eta_x,\eta_y)=\mathcal{H}'(\eta_b)+H'(\eta_u)
\end{equation}
\begin{equation}
		H'(\eta_x,\eta_y) = u_0+\delta_u\eta_b\tau_z\sigma_z+\delta_{\mu_s}\eta_u\tau_z\sigma_0+ (B_0+\delta_B\eta_b\tau_z\sigma_0+\delta_{D'}\eta_u\tau_z\sigma_z)(\cos k_x-\cos k_y)
	\label{eq:S18}
\end{equation}
Here, $\mathcal{H}'(\eta_b)$ is the term in Eq.~\eqref{eq:S15} with $\eta_b=(\eta_x+\eta_y)/2$, while $H'(\eta_u)$ is from Eq.~\eqref{eq:S16} with $\eta_u=(\eta_x-\eta_y)/2$. $u_0$ and $B_0$ are the initial parameters before strain.

\section{Phase Diagram}
From Eq.~\eqref{eq:S18}, we can calculate the gap at each valley. The gap at spin-up (X) valley is
\begin{align*}
2|(u_0+\delta_u\eta_b+ \delta_{\mu_s}\eta_u)- 2(B_0+\delta_B\eta_b+ \delta_{D'}\eta_u)|
\end{align*}
and the gap at spin-down (Y) valley is
\begin{align*}
2|(u_0+\delta_u\eta_b- \delta_{\mu_s}\eta_u)- 2(B_0+\delta_B\eta_b- \delta_{D'}\eta_u)|.
\end{align*}
From part III, we know that the condition for a topological phase transition is given by:
\begin{align*}
	\left|\frac{u_{\uparrow}}{B_{\uparrow}}\right|<2\Rightarrow\mathcal{C}_{\uparrow}=\mathrm{sgn}\left(tB_{\uparrow} \lambda\right)
\end{align*}
\begin{align*}
	\left|\frac{u_{\downarrow}}{B_{\downarrow}}\right|<2\Rightarrow\mathcal{C}_{\downarrow}=-\mathrm{sgn}\left(tB_{\downarrow} \lambda\right)
\end{align*}
Thus, we obtain the general critical condition for a topological phase transition under arbitrary strain:
\begin{equation}
\frac{\left|2B_0+\delta_B\left(\eta_x+\eta_y\right)+\delta_{D'}\left(\eta_x-\eta_y\right)\right|}{\left|2u_0+\delta_u\left(\eta_x+\eta_y\right)+\delta_{\mu_s}\left(\eta_x-\eta_y\right)\right|}>\frac{1}{2}
	\label{eq:S19}
\end{equation}
\begin{equation}
\frac{\left|2B_0+\delta_B\left(\eta_x+\eta_y\right)-\delta_{D'}\left(\eta_x-\eta_y\right)\right|}{\left|2u_0+\delta_u\left(\eta_x+\eta_y\right)-\delta_{\mu_s}\left(\eta_x-\eta_y\right)\right|}>\frac{1}{2}
	\label{eq:S20}
\end{equation}
Here, Eq.~\eqref{eq:S19} is for $\mathcal{C}_{\uparrow}=\mathrm{sgn}(tB\lambda)$ and Eq.~\eqref{eq:S20} is for $\mathcal{C}_{\downarrow}=-\mathrm{sgn}(tB\lambda)$ (In this case, we assume that the strain will not change the sign of B, i.e. $\mathrm{sgn}(B)=\mathrm{sgn}\left(B_{up}\right)=\mathrm{sgn}(B_{dn})$). 

By properly adjusting the orientation of the magnetic moments, the spin-up valley can always be positioned at the X point (if it is initially located at Y, a $90^\circ$ rotation of the entire system brings it to X). In this way, from ~\eqref{eq:S5}, the absolute value in Eq.~\eqref{eq:S19} can be removed, yielding two distinct critical curves:
\begin{equation}
    \eta_y=s_{r}(\eta_x-\eta_0)+\eta_0,\eta_y=s_{b}(\eta_x-\eta_0)+\eta_0.
    \label{twoline}
\end{equation}
\begin{align*}
s_{r}=\frac{1}{s_{b}}=-\frac{(\delta_u-2\delta_B)+(\delta_{\mu_s}-2\delta_{D'})}{(\delta_u-2\delta_B)-(\delta_{\mu_s}-2\delta_{D'})},\eta_0=\frac{u_0-2B_0}{2\delta_B-\delta_u}. 
\end{align*}
Here, $k$ and $1/k$ are the slopes of the two critical curves, and $(\eta_0,\eta_0)$ denotes their intersection point.
Therefore, starting from the trivial case with $u_0>2B_0$, the system can undergo transitions into any of the other three phases by strain.
If, upon increasing strain from the origin, the trajectory crosses the first curve but not the second, the system enters a QAH phase with a spin-up polarized edge state. Conversely, crossing only the other curve yields a QAH phase with spin-down edge polarization, carrying the opposite Chern number. If both curves are crossed, the system instead evolves into the Type-II QSH phase.

\section{DFT Calculation}

\begin{figure}
	\centering 
	\includegraphics[width=0.8\textwidth]{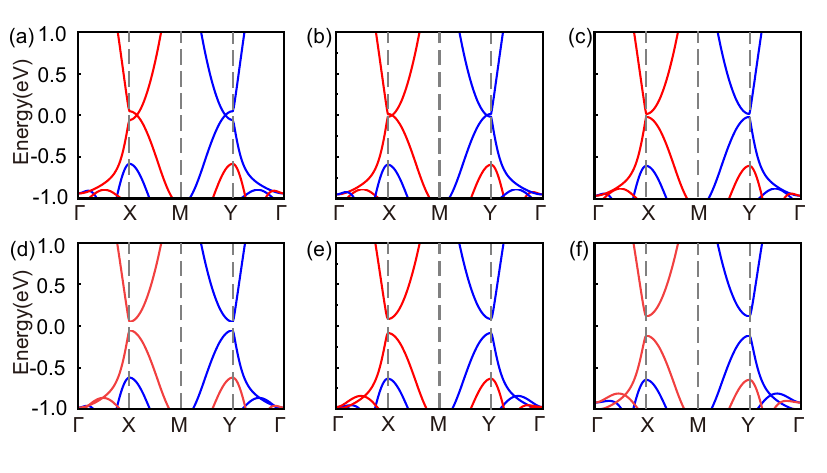}%
	\caption{\label{fig:S7} 
		Band structure (without SOC) under different lattice constants ($a_x=a_y=a$). Panels (a)–(f) correspond to lattice constants ranging from $3.30\AA$ to $3.40\AA$. The red line represents the spin-up and black represents spin-down.}
\end{figure}

\begin{figure}
	\centering 
	\includegraphics[width=0.8\textwidth]{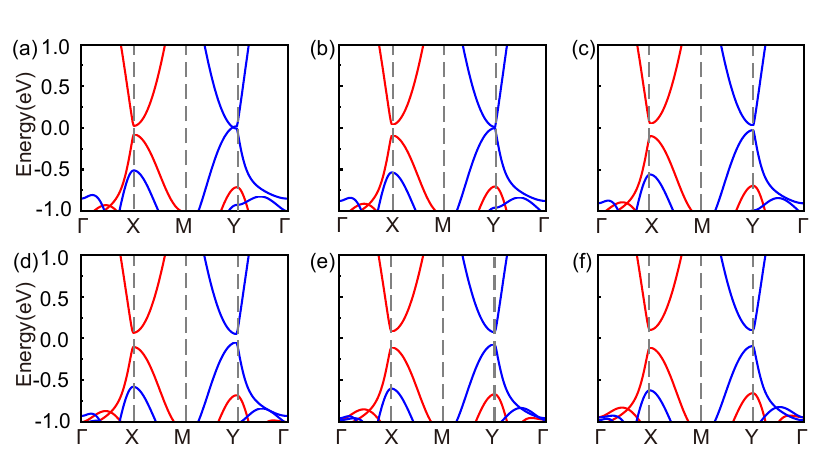}%
	\caption{\label{fig:S8} 
		Band structure (without SOC) under different $a_x$ while $a_y=3.40$\AA. Panels (a)–(f) correspond to $a_x$ ranging from $3.28$\AA to $3.38$\AA. The red line represents the spin-up and black represents spin-down.}
\end{figure}

\begin{figure}
	\centering 
	\includegraphics[width=0.8\textwidth]{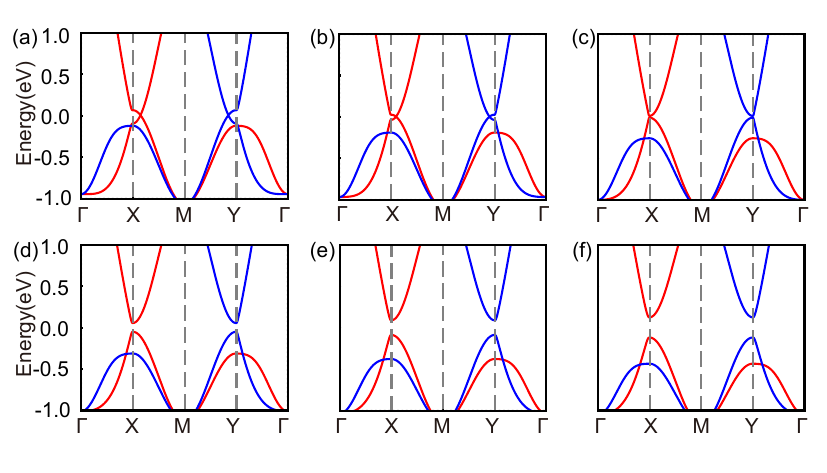}%
	\caption{\label{fig:S9} 
		The band structures calculated using the fitted parameters from Table~\ref{tab:SI} at different lattice constants ($a_x=a_y=a$) exhibit essentially identical properties near the valleys with DFT calculation. Panels (a)–(f) correspond to lattice constants ranging from $3.30$\AA to $3.40$\AA.}
\end{figure}

\begin{figure}
	\centering 
	\includegraphics[width=0.8\textwidth]{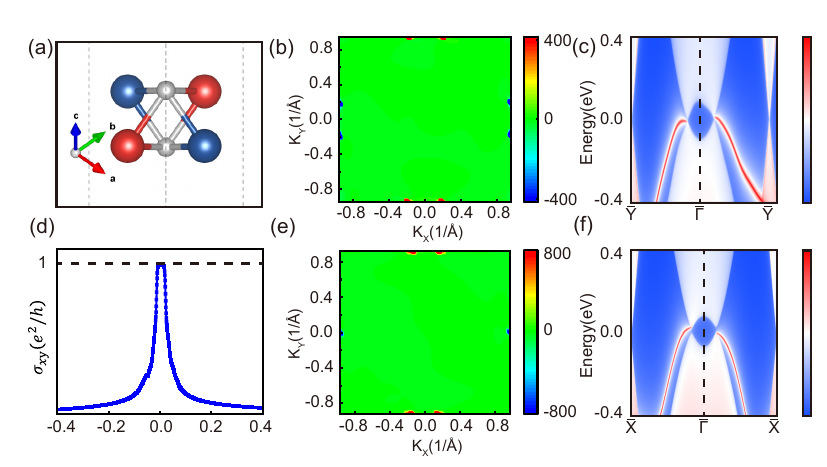}%
	\caption{\label{fig:S10} 
		(a) Schematic diagram of monolayer CrO. (b) Berry curvature and (c) edge state of CrO monolayer at $a_x=3.30$\AA, $a_y=3.30$\AA. (e) Berry curvature and (f) edge state of CrO monolayer at $a_x=3.24$\AA, $a_y=3.40$\AA. (d) Anomalous Hall conductivity vs chemical potential at this point. To better visualize the edge states, in panel (c) we slightly enlarge part of the plotted path, which reveals an additional edge state at Y and confirms the type-II QSH phase. Furthermore, in panel (f) we modify the path so that the edge state originally located at Y is shifted to the center, making it more evident.}
\end{figure}

First-principles calculations have been performed using the Vienna ab initio simulation package (VASP), which is based on density functional theory (DFT)\cite{SM1,SM2,SM3}, The Perdew-Burke-Ernzerhof functional at the level of generalized gradient approximation (GGA) is used to deal with the exchange correlation interactions\cite{SM4}. The correlation effect for the Cr 3d electrons is treated by the DFT+U method\cite{SM5,SM6}, with $\mathrm{U_{Cr}=3.55eV}$. The plane-wave basis is used with a cutoff energy of 520eV. The Brillouin zone is sampled with a $\Gamma$-centered $k$ mesh of size $15\times15\times1$. The crystal structure is fully optimized until the force applied on each atom is less than 0.001eV/\AA. The energy convergence criterion is set to be ${10}^{-8}$eV. A vacuum layer of $20$\AA\ is used to avoid residual interactions between neighboring layers. The Berry curvature of occupied bands is calculated by the WANNIER TOOLS software package\cite{SM7} and VASPBERRY software package\cite{SM8}. The topological edge states are calculated by the WANNIER90 software package\cite{SM9} with the iterative Green’s function approach\cite{SM10}.

The stability and magnetic properties of CrO can be referenced from Ref.\cite{SM11}. In our work, we primarily focus on its band structure and whether the evolution of its topological phase with varying lattice constants agrees with our earlier model analysis. As shown in Fig.\ref{fig:S7}, the band structures under different lattice constants reveal that energy valleys emerge at both the X and Y points. With decreasing lattice constant, the bandgap gradually shrinks. When the lattice constant falls below $3.33$\AA, double degenerate points appear with linearly dispersion, signaling a transition from an insulating phase to a semimetal, which corresponds to a transition from a topologically trivial phase to a QSH phase.

Fig.~\ref{fig:S8} further displays the evolution of the band structure as a function of $a_x$, with $a_y$ fixed at $3.40$\AA. It shows that when $a_x<3.29$\AA, a gap closing occurs in the spin-down sector at the Y valley, indicating a transition from a topologically trivial state to a QAH phase.

The Berry curvature and edge states also show the untrivial phase in Fig.~\ref{fig:S10}, confirm the predictions of our theoretical model: the topological phase transitions (from trivial to QSH or QAH) can be realized in CrO by tuning its lattice constants, demonstrating the practical relevance of our model in real material systems.

\begin{table}[htbp]
	\centering
	\caption{The fitted parameters when $a_x = a_y$.}
	\label{tab:SI}
	\begin{tabular}{ccccccccc}
		\toprule
		$a_x$ & $A$ & $B$ & $C$ & $D$ & $\lambda$ & $\mu$ & $t$ & $u$ \\
		\midrule
		3.28 & 0.147787 & -0.39072 & 0.35052  & 0.390718 & 0 & 0.676027 & 1.893019 & -0.65088 \\
		3.30 & 0.144117 & -0.37527 & 0.328334 & 0.375265 & 0 & 0.643942 & 1.875051 & -0.67069 \\
		3.32 & 0.140485 & -0.36036 & 0.306404 & 0.360352 & 0 & 0.609416 & 1.855523 & -0.68986 \\
		3.34 & 0.137229 & -0.34756 & 0.285603 & 0.347560 & 0 & 0.573530 & 1.848751 & -0.70630 \\
		3.36 & 0.135049 & -0.33463 & 0.269236 & 0.334628 & 0 & 0.540948 & 1.821552 & -0.72241 \\
		3.38 & 0.130524 & -0.32387 & 0.251323 & 0.323866 & 0 & 0.505085 & 1.800626 & -0.73642 \\
		3.40 & 0.126751 & -0.31389 & 0.234977 & 0.313891 & 0 & 0.472278 & 1.776148 & -0.74971 \\
		\bottomrule
	\end{tabular}
\end{table}

\begin{table}[htbp]
	\centering
	\caption{The fitted parameters when $a_x \neq a_y$ and $a_y=3.40$\,\AA.}
	\label{tab:S2}
	\begin{tabular}{ccccccccc}
	\toprule
	$a_x$ & $A$ & $B$ & $C$ & $D$ & $\lambda$ & $\mu$ & $t$ & $u$ \\
	\midrule
			3.26up & 0.083802 & -0.31595 & 0.263916 & 0.315947 & 0 & 0.502511 & 1.816002 & -0.67554 \\
			3.26dn & 0.196097 & -0.38300 & 0.324700 & 0.382999 & 0 & 0.656201 & 1.848852 & -0.70561 \\
			3.28up & 0.092418 & -0.31613 & 0.262714 & 0.316123 & 0 & 0.499803 & 1.803907 & -0.68939 \\
			3.28dn & 0.186594 & -0.37159 & 0.310750 & 0.371590 & 0 & 0.627663 & 1.840279 & -0.71321 \\
			3.30up & 0.100794 & -0.31623 & 0.261234 & 0.316229 & 0 & 0.498563 & 1.791932 & -0.70210 \\
			3.30dn & 0.176912 & -0.36921 & 0.296000 & 0.369204 & 0 & 0.599918 & 1.842262 & -0.71133 \\
			3.32up & 0.109007 & -0.31628 & 0.259597 & 0.316276 & 0 & 0.498665 & 1.779480 & -0.71395 \\
			3.32dn & 0.167282 & -0.35076 & 0.283370 & 0.350759 & 0 & 0.573581 & 1.832207 & -0.72592 \\
			3.34up & 0.117131 & -0.31621 & 0.257776 & 0.316203 & 0 & 0.499724 & 1.766461 & -0.72476 \\
			3.34dn & 0.157451 & -0.34074 & 0.271040 & 0.340743 & 0 & 0.548096 & 1.819415 & -0.73213 \\
			3.36up & 0.125057 & -0.31578 & 0.255606 & 0.315780 & 0 & 0.501153 & 1.753754 & -0.73460 \\
			3.36dn & 0.149213 & -0.33134 & 0.260740 & 0.331336 & 0 & 0.526879 & 1.799510 & -0.73809 \\
			3.38up & 0.133263 & -0.31473 & 0.252805 & 0.314726 & 0 & 0.501958 & 1.740731 & -0.74275 \\
			3.38dn & 0.144672 & -0.32229 & 0.254550 & 0.322283 & 0 & 0.513555 & 1.766163 & -0.74405 \\
		\bottomrule
\end{tabular}
\end{table}

Tables~\ref{tab:SI} and ~\ref{tab:S2} summarize the parameters obtained by fitting the model to the band structure data. Based on these fitted parameters, we can extract the key quantities relevant to constructing the topological phase diagram. The procedure is as follows:

Under biaxial strain $\eta_s$, from Eq.~\eqref{eq:S16}, we get: $u_{\uparrow/\downarrow}=u_0+\delta_u\eta_b$ and $B_{\uparrow/\downarrow}=B_0+\delta_B\eta_b$.
Under initial conditions, $a_0=3.40$\AA, thus, $u_0=u\left(a_0\right)=u(3.40)$, $B_0=B\left(a_0\right)=B(3.40)$, then: 
\begin{align*}
	\delta_u=\frac{\partial u}{\partial\eta}=\frac{\partial u}{\partial a}\frac{\partial a}{\partial\eta}.
\end{align*}
Since $\eta=\left(a-a_0\right)/a_0$, we obtain:
\begin{align*}
	\delta_u=a_0\frac{\partial u}{\partial a}.
\end{align*}
We use the data at lattice constants 3.40\AA\ and 3.38\AA\ in table~\ref{tab:SI} to obtain the partial derivatives.
\begin{align*}
	\delta_u=a_0\frac{\partial u}{\partial a}=a_0\left(\frac{u\left(3.40\right)-u(3.38)}{3.40-3.38}\right)=-2.2593
\end{align*}
Similarly, for $\delta_B$:
\begin{align*}
	\delta_B=a_0\frac{\partial B}{\partial a}=a_0\left(\frac{B\left(3.40\right)-B(3.38)}{3.40-3.38}\right)=1.6966
\end{align*}
From the definition of $\eta_0$ 
\begin{align*}
	\eta_0=\frac{u_0-2B_0}{2\delta_B-\delta_u}=-0.02157
\end{align*}
The corresponding lattice constant is $a=a_0\left(1+\eta_0\right)=3.327$\AA.
Under the strain $\eta_x$ and $\eta_y$, the parameter will change as (from~\eqref{eq:S19} and~\eqref{eq:S20}):
\begin{align*}
	u_{\uparrow/\downarrow}=u_0+\delta_u(\eta_x+\eta_y)/2\pm\delta_{\mu_s}(\eta_x-\eta_y)/2\\
	B_{\uparrow/\downarrow}=B_0+\delta_B(\eta_x+\eta_y)/2\pm\delta_{D'}(\eta_x-\eta_y)/2
\end{align*}
Under initial conditions, $a_0=3.40$\AA, we get:
\begin{align*}
	\delta_{\mu_s}=a_0\left(\frac{\partial u_{\uparrow}}{\partial a_x}-\frac{\partial u_{\downarrow}}{\partial a_x}\right)\\
	\delta_{D'}=a_0\left(\frac{\partial B_{\uparrow}}{\partial a_x}-\frac{\partial B_{\downarrow}}{\partial a_x}\right)
\end{align*}
Similarly, using the data at lattice constants $3.40$\AA\ and $3.38$\AA\ in table \ref{tab:S2} to obtain the partial derivatives.
\begin{align*}
	\delta_{\mu_s}=a_0\left(\frac{u_{\uparrow}\left(3.40\right)-u_{\uparrow}(3.38)}{3.40-3.38}-\frac{u_{\downarrow}\left(3.40\right)-u_{\downarrow}(3.38)}{3.40-3.38}\right)=0.2957\\
	\delta_{D'}=a_0\left(\frac{B_{\uparrow}\left(3.40\right)-B_{\uparrow}(3.38)}{3.40-3.38}-\frac{B_{\downarrow}\left(3.40\right)-B_{\downarrow}(3.38)}{3.40-3.38}\right)=4.0928
\end{align*}

\bibliography{SM1.bib}